\documentclass[prd,aps,floats,showkeys,nofootinbib,notitlepage]{revtex4}
\usepackage{amsmath}
\usepackage{amssymb}
\usepackage{amsbsy}
\usepackage{float}
\restylefloat{table}
\usepackage{graphicx}
\usepackage{epsfig}
\usepackage{epstopdf}
\usepackage{subfig}
\usepackage[colorlinks=true,linkcolor=red]{hyperref}
\usepackage{listings} 
\usepackage{xcolor} 
\usepackage{hyperref}
\usepackage{tabularx}
\usepackage{multirow}
\usepackage{enumerate}
\usepackage{hyperref}

\floatstyle{plaintop}
\restylefloat{table}

\newcolumntype{Y}{>{\centering\arraybackslash}X} 

\newcommand{\matrixHorizon}[1]{\ensuremath{\overset{\xrightarrow[\hphantom{#1}]{\text{flipping horizontally}}}{#1}}} 
\newcommand{\matrixVertical}[1]{\ensuremath{\left\downarrow\vphantom{#1}\right.{#1}}} 

\definecolor{ForestGreen}{RGB}{34,139,34}
\definecolor{OliveGreen}{RGB}{72.9,72.2,42.4}
\DeclareCaptionFormat{listing}{\rule{\dimexpr\textwidth}{0.4pt}\par\vskip1pt#1#2#3}
\begin{document}
	\title{A New Algorithm to determine Adomian Polynomials for nonlinear polynomial functions}
	\author{Mithun Bairagi}
	\email{bairagirasulpur@gmail.com}
	\affiliation{Mangal Chandi High School, Khosalpur, 722206, Patrasayer, Bankura, West Bengal, India}

	\begin{abstract}
	We present a new algorithm by which the Adomian polynomials can be determined for scalar-valued nonlinear polynomial functional in a Hilbert space. This algorithm calculates the Adomian polynomials without the complicated operations such as parametrization, expansion, regrouping, differentiation, etc. The algorithm involves only some matrix operations. Because of the simplicity in the mathematical operations, the new algorithm is faster and more efficient than the other algorithms previously reported in the literature. We also implement the algorithm in the MATHEMATICA code. The computing speed and efficiency of the new algorithm are compared with some other algorithms in the one-dimensional case.
	
	\end{abstract}
	\maketitle

	\section{introduction} 
	The Adomian Decomposition Method (ADM) \cite{adm,adm1,adm2,adm3,adm4} has gained huge attention in different fields of science and engineering for solving nonlinear functional equations.
In practice, many nonlinear problems do not admit exact solutions, and in most cases, we have to find approximate solutions by employing numerical or analytical approximation techniques. 
The ADM is a reliable technique for solving wide classes of nonlinear systems, including ordinary differential, partial differential, integro-differential, algebraic, differential-algebraic, non-integer-order differential, integral equations, and so on \cite{Mavoungou,Ngarhasta,admapp,admapp1,admapp2,admapp3,admapp4}). This technique can provide an analytical approximation to the exact solutions in the series form that converge very rapidly \cite{admconver,admconver1,admconver2}.
The Adomian decomposition method coupled with the Laplace transform, develops a powerful method called the Laplace Adomian decomposition method (LADM). LADM has also been used in numerous articles to find the numerical solution of fractional-order nonlinear differential equations, as can be seen in \cite{ladm1,ladm2,ladm3,ladm4,ladm5}.
	
Following \cite{Mavoungou,Ngarhasta,Duan}, let us recall the basic ideas of the Adomian Decomposition Method. We consider a nonlinear ODE in order $p$ with independent variable $x$ (real and scalar) and dependent variable $u$ in the general form \cite{Mavoungou,Ngarhasta}
\begin{equation}
	\mathcal{F}u=g(x),
\end{equation}
where $\mathcal{F}$ is the nonlinear operator from a Hilbert space $H$ into $H$. In ADM, $\mathcal{F}$ is assumed to be decomposed into
\begin{equation}\label{nlode}
	Lu+Ru+Nu=g(x),
\end{equation}
where $L$ is the highest-order linear differential operator $L[.]=\frac{d^p}{dx^p}[.]$ which is assumed to be invertible, $R$ is a linear differential operator containing the linear derivatives of less order than $L$, $N$ is a nonlinear operator containing all other nonlinear terms, $g(x)\in H$ is a given analytic function. Here we should note that the choice of the operator $L$ is not generally unique \cite{adomian95,wazwaz02,wazwaz06}. For example, in \cite{wazwaz02}, A. Wazwaz has chosen the linear differential operator $L[.]$ as $L[.]=x^{-2}\frac{d}{dx}\left(x^2\frac{d}{dx}\right)$ for the Lane-Emden equation.
It is also notable that $u$ is a scalar function of real variable $x$ in Eq. \eqref{nlode}. For a system of differential equations, $u$ will be a vector-valued function. However, in this paper, our studies are restricted to single ODE where $u$ is a scalar-valued function.
The principle step of the decomposition method is to suppose a series solution defined by
	\begin{equation}\label{seriesSolu}
		u=\sum_{i=0}^{\infty}u_i,
	\end{equation}
and then the ADM scheme corresponding to the functional equation \eqref{nlode} converges rapidly to $u\in H$ which is the unique solution to the functional equation \cite{Mavoungou,Bougoffa}.
Equation \eqref{seriesSolu} decomposes the nonlinear term $Nu$ into an infinite series 
	\begin{equation}\label{Nu}
		Nu=\sum_{i=0}^{\infty}A_i,
	\end{equation}
	where $A_i$ are the so-called Adomian polynomials which depend on the solution components $u_0,u_1,\ldots,u_i$. For a given nonlinear functional $Nu=F(u)$ ($F(u)$ is assumed to be an analytic function of variable $u$ in Hilbert space $H$), the Adomian polynomials are determined by the following definitional
	formula introduced by G. Adomian \cite{adm,adm1,adm2,Duan3}:
	\begin{equation}\label{admdef}
		A_M=\left.\frac{1}{M!}\frac{d^M }{d \lambda^M}{F\left(\sum_{k=0}^{\infty}u_k\lambda^k\right)}\right|_{\lambda=0},\;\; \;\;M=0,1,2,\ldots,
	\end{equation}
	where the analytic parameter $\lambda$ is simply a grouping parameter. An important property of Adomian polynomial $A_M$ is that it depends by construction only on the solution components $(u_0,u_1,\ldots,u_M)$ and does not depend on higher-order solution components $u_k$ with $k>M$ \cite{Wazwaz,Azreg}. Therefore, the higher-order terms for $k>M$ do not contribute in summation in Eq. \eqref{admdef}. 
	
	Main step of ADM is to determine the Adomian polynomials of the nonlinear term $Nu$.
	Using the definitional formula \eqref{admdef} it is difficult to calculate higher-order Adomian terms due to the complexity in calculations of higher-order derivatives. Later, many authors have developed several convenient algorithms for fast generation of the one-variable and the multi-variable Adomian polynomials. Adomian and Rach \cite{AR} produced a recurrence rule that provides a systematic computational procedure to determine Adomian polynomials. Later, Rach in his paper \cite{Rach} established simple symmetry rules (which is called Rach's rule) in Adomian and Rach's algorithms, by which Adomian’s polynomials can be determined quickly to higher orders. Using the algorithm presented by Wazwaz in \cite{Wazwaz}, we need to collect the terms from the expansion, which takes a large computational time for higher orders. Applying the algorithm in \cite{Biazar}, we require to compute the derivative after substitution in a recurrence relation between the Adomian polynomials. Recently in \cite{Agom}, the authors modified the formula \eqref{admdef} to determine the Adomian polynomials for nonlinear polynomial functionals. 
	In \cite{Duan,Duan1}, Duan has developed more efficient and fast recurrence algorithms for the rapid generation of the Adomian polynomials for one-variable (which is the one-dimensional case in our studies) and multi-variable cases. Duan’s Corollary 1 algorithm \cite{Duan} (called index recurrence algorithm) and Duan’s Corollary 3 algorithm \cite{Duan1} do not involve the differentiation operator in determining the reduced polynomials in one dimension. We only require the operations of addition and multiplication, which make these algorithms faster and more efficient techniques. 
	
	In this work, we have presented a new algorithm for fastest computations of Adomian polynomials for scalar-valued nonlinear polynomial functional (with index as positive integers) in a Hilbert space $H$ with the help of matrix formulations rather than recurrence processes. Our proposed algorithm does not require complex mathematical operations such as parametrization, expansion, regrouping, and differentiation. In this algorithm, the higher-order Adomian polynomials can be determined through few matrix operations, making it faster and more efficient than the other existing algorithms in the literature. We have generalized the new algorithm in two dimensions where the solution $u$ depends on two-state variables such as $t,x$.

	The paper is organized as follows: In Sec. \ref{algo} we present our algorithm to determine Adomian polynomials for nonlinear polynomial functional. In Sec. \ref{comp}, we apply our algorithms to the polynomial functions, and the computation times are compared with some other popular algorithms previously reported in the literature. In Sec. \ref{con}, we discuss our results and make some conclusions on our works. We list the MATHEMATICA code for the new algorithms in Listing \ref{onedOur} for one-dimensional case and in Listing \ref{twodOur} for two-dimensional case in Appendix: \ref{oned}, \ref{twod} respectively. We have also listed the MATHEMATICA code for some other algorithms which are Duan’s Corollary 1 algorithm \cite{Duan} and Duan’s Corollary 3 algorithm \cite{Duan1,Duan2} with the one-dimensional case in Listings \ref{onedDuan}, \ref{onedDuan1} in Appendix: \ref{oned}. 

\section{description of our proposed algorithm}\label{algo}
	In this section, we have described a new algorithm for calculating the Adomian polynomials. This algorithm is only applicable for scalar-valued nonlinear polynomial functional (with index as positive integers) in a Hilbert space $H$ for the two-dimensional case. In order to increase the calculating efficiency in this algorithm, all the mathematical operations are performed in the matrix forms. 
	
	Let us now consider a nonlinear polynomial functional $F$ depends on two different functions $u$ and $v$ in $H$. 
	The functions $u$ and $v$ can be expanded into the following two-dimensional series
	\begin{equation}\label{uvSeries}
		u=\sum_{i=0}^{\infty}\sum_{j=0}^{\infty}u_{ij}\;\;\;\text{and}\;\;\;v=\sum_{i=0}^{\infty}\sum_{j=0}^{\infty}v_{ij}\;.
	\end{equation}
	To illustrate our algorithm, we take the nonlinearity $F$ in the simple form 
	\begin{equation}\label{nlF}
		F=uv.
	\end{equation}
	And this nonlinear function can be decomposed by a series
	\begin{equation}
		F=\sum_{i=0}^{\infty}\sum_{j=0}^{\infty}A_{ij},
	\end{equation}
    where $A_{ij}$ are called Adomian polynomials of the components $u_{ij},v_{ij}\;(i=0,1,\ldots,j=0,1,\ldots)$.
	Now, we divide the algorithm into six main steps (labeled from Step-1 to Step-6), and to illustrate each step, we have used the nonlinear polynomial function \eqref{nlF}.

	\begin{itemize}
		\item[]\textbf{Step-1} (Express the functions $u$ and $v$ in the matrix forms): In this step, the functions $u$ and $v$ are expressed in the matrix forms. For computations in computer, we truncate the infinite series \eqref{uvSeries} up to the finite terms $i=m,j=n$. We can increase the accuracy in our results by increasing the values of $m,n$ as far as possible. The functions $u,v$ in the Eq. \eqref{uvSeries} can be expressed by $(m+1)\times (n+1)$ matrices
		\begin{equation}\label{soluMatrix}
			U=
			\begin{pmatrix}
				u_{00}&  u_{01}& \ldots& u_{0l}& \ldots& u_{0n}\\ 
				\vdots&	 \vdots& \ldots& \vdots& \ldots& \vdots\\
				u_{k0}&  u_{k1}& \ldots& u_{kl}& \ldots& u_{kn}\\ 
				\vdots&	 \vdots& \ldots& \vdots& \ldots& \vdots\\
				u_{m0}&  u_{m1}& \ldots& u_{ml}& \ldots& u_{mn}
			\end{pmatrix} \;\;\text{and}\;\; 
			V=
			\begin{pmatrix}
				v_{00}&  v_{01}& \ldots& v_{0l}& \ldots& v_{0n}\\ 
				\vdots&	 \vdots& \ldots& \vdots& \ldots& \vdots\\
				v_{k0}&  v_{k1}& \ldots& v_{kl}& \ldots& v_{kn}\\ 
				\vdots&	 \vdots& \ldots& \vdots& \ldots& \vdots\\
				v_{m0}&  v_{m1}& \ldots& v_{ml}& \ldots& v_{mn}
			\end{pmatrix}. 
		\end{equation}
	
		\item[]\textbf{Step-2} (Extracting the submatrices from the matrices $U$ and $V$): The Adomian polynomials corresponding to any matrix elements (let the matrix elements $u_{kl},v_{kl}$ located at row $k+1$, column $l+1$) in Eq. \eqref{soluMatrix}, depend on the other matrix elements whose row number ($r$) and column number ($c$) are less than or equal to $k+1$ and $l+1$ respectively, but do not depend on the matrix elements located at $r>k+1$ and $c>l+1$. In order to calculate the Adomian polynomials for the elements $u_{kl}$ and $v_{kl}$ in $U$ and $V$, we extract the submatrices formed by the elements with rows $r\leq k+1$ and columns $c\leq l+1$ of the matrices $U$ and $V$ in Eq. \eqref{soluMatrix}. These submatrices are given by 
		\begin{equation}\label{subMatrix}
			U[{0,1,\ldots,k;0,1,\ldots,l}]=
			\begin{pmatrix}
				u_{00}&  u_{01}& \ldots& u_{0l}\\ 
				\vdots&	 \vdots& \ldots& \vdots\\
				u_{k0}&  u_{k1}& \ldots& u_{kl}\\ 				
			\end{pmatrix} \;\;\text{and}\;\; 
			V[{0,1,\ldots,k;0,1,\ldots,l}]=
			\begin{pmatrix}
				v_{00}&  v_{01}& \ldots& v_{0l}\\ 
				\vdots&	 \vdots& \ldots& \vdots\\
				v_{k0}&  v_{k1}& \ldots& v_{kl}			
			\end{pmatrix}. 
		\end{equation}
	
	\item[] \textbf{Step-3} (Flipping the submatrix): In this step, all the matrix elements of any one of the submatrices in Eq. \eqref{subMatrix} are flipped horizontally and then vertically or vice versa. Here we perform the flipping operation on the submatrix $V[{0,1,\ldots,k;0,1,\ldots,l}]$. The flipping operation along horizontal axis can be shown in the following way
	\begin{equation}
		\matrixHorizon{	\begin{pmatrix}
			v_{00}&  v_{01}& \ldots& v_{0l}\\ 
			\vdots&	 \vdots& \ldots& \vdots\\
			v_{k0}&  v_{k1}& \ldots& v_{kl}			
		\end{pmatrix}}\longrightarrow
		\begin{pmatrix}
		v_{0l}&  v_{0l-1}& \ldots& 	v_{00}\\ 
			\vdots&	 \vdots& \ldots& \vdots\\
			v_{kl}&  v_{kl-1}& \ldots& v_{k0}			
		\end{pmatrix}=V[{0,1,\ldots,k;l,l-1,\ldots,0}].
	\end{equation}
	Then, the flipping operation along vertical axis is performed on the above flipped submatrix, which can be shown as 
	\begin{equation}
		\text{\tiny flipping vertically}\matrixVertical{	\begin{pmatrix}
				v_{0l}&  v_{0l-1}& \ldots& 	v_{00}\\ 
				\vdots&	 \vdots& \ldots& \vdots\\
				v_{kl}&  v_{kl-1}& \ldots& v_{k0}			
		\end{pmatrix}}\longrightarrow
		\begin{pmatrix}
			v_{kl}&  v_{kl-1}& \ldots& 	v_{k0}\\ 
			\vdots&	 \vdots& \ldots& \vdots\\
			v_{0l}&  v_{0l-1}& \ldots& v_{00}			
		\end{pmatrix}=V[{k,k-1,\ldots,0;l,l-1,\ldots,0}].
	\end{equation}

	\item[]\textbf{Step-4} (Element-wise matrices multiplication): In the element-wise multiplication (also known as the Hadamard product), each element $i,j$ in the two matrices are multiplied together. We perform the element-wise multiplication between two matrices $U[{0,1,\ldots,k;0,1,\ldots,l}]$ and $V[{k,k-1,\ldots,0;l,l-1,\ldots,0}]$, given by
	\begin{equation} 
		U[{0,1,\ldots,k;0,1,\ldots,l}]\circ V[{k,k-1,\ldots,0;l,l-1,\ldots,0}]=W[{0,1,\ldots,k;0,1,\ldots,l}]
	\end{equation}
	and in the matrix notation the above equation can be expressed by
	\begin{equation}
		\begin{pmatrix}
			u_{00}&  u_{01}& \ldots& u_{0l}\\ 
			\vdots&	 \vdots& \ldots& \vdots\\
			u_{k0}&  u_{k1}& \ldots& u_{kl}\\ 				
		\end{pmatrix}\circ \begin{pmatrix}
								v_{kl}&  v_{kl-1}& \ldots& 	v_{k0}\\ 
								\vdots&	 \vdots& \ldots& \vdots\\
								v_{0l}&  v_{0l-1}& \ldots& v_{00}			
							\end{pmatrix}
		 = 	\begin{pmatrix}
			 	u_{00} v_{kl}&  u_{01} v_{kl-1}& \ldots& u_{0l}v_{k0}\\ 
			 	\vdots&	 \vdots& \ldots& \vdots\\
			 	u_{k0}v_{0l}&  u_{k1}v_{0l-1}& \ldots& u_{kl}v_{00}			
		 	\end{pmatrix}.
	\end{equation}
	Here the symbol $\circ$ denotes the  element-wise multiplication between two matrices.
	\item[]\textbf{Step-5} (Summation over matrix elements): In this step, we take summation over all the elements of the matrix $W[{0,1,\ldots,k;0,1,\ldots,l}]$ and this summation is
	\begin{equation}\label{Akl}
		A_{kl}=\sum_{i=0}^{k}\sum_{j=0}^{l}W_{ij} = u_{00} v_{kl}+  u_{01} v_{kl-1}+\ldots+u_{kl}v_{00}.
	\end{equation}
	Here $A_{kl}$ is the Adomian polynomial for the two matrix elements $u_{kl},v_{kl}$. In the Adomian polynomial $A_{kl}$, notably, the sum of the first index at subscripts of the components of $u,v$ in each term in $A_{kl}$ are same. Similarly, the sum of the second index of the components of $u,v$ in each term in $A_{kl}$ are also same (here for the first index, the sum is $k$ and for the second index, the sum is $l$), which obey the important property of the Adomian polynomial given in \cite{Wazwaz}.   
	
	\item[]\textbf{Step-6} (Constructing Adomian matrix): Repeating the previous steps from Step-1 to Step-5, the Adomian polynomials corresponding to each matrices elements in Eq. \eqref{soluMatrix} are determined. All the calculated Adomian polynomials are stored in a matrix and can be expressed by
	\begin{equation}\label{adomianM}
		A=
		\begin{pmatrix}
			A_{00}&  A_{01}& \ldots& A_{0l}& \ldots& A_{0n}\\ 
			\vdots&	 \vdots& \ldots& \vdots& \ldots& \vdots\\
			A_{k0}&  A_{k1}& \ldots& A_{kl}& \ldots& A_{kn}\\ 
			\vdots&	 \vdots& \ldots& \vdots& \ldots& \vdots\\
			A_{m0}&  A_{m1}& \ldots& A_{ml}& \ldots& A_{mn}
		\end{pmatrix}.
	\end{equation}
	We call the matrix $A$ in \eqref{adomianM} as Adomian matrix for the given polynomial nonlinearity \eqref{nlF}.
	
\end{itemize} 
 
We present the pseudo-code for the algorithms described in Step-1 to Step-6 in Listing \ref{alg1} which compute the Adomian matrix of Eq. \eqref{nlF}.
Here, it is worthwhile to note how a few simple matrix operations in Step-1 to Step-6 generate the Adomian polynomials of Eq. \eqref{nlF}. 
It is clear from Step-1 to Step-6 that only $4(m+1)(n+1)-(m+n+2)$ number of matrix operations ($2(m+1)(n+1)-(m+n+2)$ number of flippings, $(m+1)(n+1)$ number of element-wise matrices multiplications and $(m+1)(n+1)$ number of matrix summations) are required to compute the Adomian matrix of Eq. \eqref{nlF} with $i=m,j=n$ in Eq. \eqref{uvSeries}.  
This simplicity in mathematical operations enhances the computing efficiency of this algorithm.

\begin{lstlisting}[numbers=left,keywordstyle=\color{black}\bfseries,	keywords={,input, output, function, for, to, do, end, return, },mathescape=true,caption={Computation of Adomian matrix $A$ of Eq. \eqref{nlF} in pseudo-code.}, label={alg1}]
input: Functions $u$ and $v$ of Eq. $\eqref{nlF}$
output: Adomian matrix $A$
function AdomianMatrix($u,v$)
	Express $u$ in matrix form $U$: $U$ $\gets$ Matrix($\sum_{i=0}^{m}\sum_{j=0}^{n}u_{ij}$)
	Express $v$ in matrix form $V$: $V$ $\gets$ Matrix($\sum_{i=0}^{m}\sum_{j=0}^{n}v_{ij}$)
	for $k\gets m$ to $k\geq 0$ do
		for $l\gets n$ to $l\geq 0$ do
			$U[{0,1,\ldots,k;0,1,\ldots,l}]$ $\gets$ the submatrix of $U$ $\text{for}$ the elements $U_{kl}$
			$V[{0,1,\ldots,k;0,1,\ldots,l}]$ $\gets$ the submatrix of $V$ $\text{for}$ the elements $V_{kl}$
			$V[{k,k-1,\ldots,0;l,l-1,\ldots,0}]$ $\gets$ $V[{0,1,\ldots,k;0,1,\ldots,l}]$ are flipped horizontally and then vertically
			Element-wise multiplication: $W[{0,1,\ldots,k;0,1,\ldots,l}]$ $\gets$ $U[{0,1,\ldots,k;0,1,\ldots,l}]\circ V[{k,k-1,\ldots,0;l,l-1,\ldots,0}]$
			$A_{kl}$ $\gets$ $\sum_{i=0}^{k}\sum_{j=0}^{l}W_{ij}$
		end for
	end for
	return A
end function    
\end{lstlisting}
\newpage
\begin{lstlisting}[numbers=left,keywordstyle=\color{black}\bfseries,	keywords={,input, output, function, for, to,do, end, return, },mathescape=true,caption={Computation of Adomian matrix $A$ of Eq. \eqref{genF} in pseudo-code.}, label={alg2}]
input: Functions $u^{(1)},u^{(2)},u^{(3)},\ldots, u^{(P-2)},u^{(P-1)},u^{(P)}$ of Eq. $\eqref{genF}$
output: Adomian matrix $A$
function AdomianMatrix2($u^{(1)},u^{(2)},u^{(3)},\ldots, u^{(P-2)},u^{(P-1)},u^{(P)}$)
	Express $u^{(1)},u^{(2)},u^{(3)},\ldots, u^{(P-2)},u^{(P-1)},u^{(P)}$ in matrix forms: $U^{(P)}$ $\gets$ Matrix($\sum_{i=0}^{m}\sum_{j=0}^{n}u_{ij}^{(P)}$)
	$A$ $\gets$ $U^{(P)}$
	for $k\gets P$ to $k\geq 2$ do
		A $\gets$ AdomianMatrix($U^{(k-1)}$,$A$)	
	end for
	return A
end function	    
\end{lstlisting}

\subsection{$F$ in general form}
Let us now consider the nonlinear polynomial functional $F$ in the following general form
\begin{equation}\label{genF}
	F=u^{(1)}u^{(2)}u^{(3)}\ldots u^{(P-2)}u^{(P-1)}u^{(P)},
\end{equation} 
where $F$ depends on $P$ number of two-dimensional functions $u^{(1)},u^{(2)},u^{(3)},\ldots ,u^{(P)}$. For $P=2$ and $u^{(1)}=u,u^{(2)}=v$, Eq. \eqref{genF} is reduced to Eq. \eqref{nlF}. The algorithms presented in the Step-1 to Step-6 also work for Eq. \eqref{genF} in the following way.
Let $U^{(1)},U^{(2)},U^{(3)},\ldots ,U^{(P)}$ are the matrix forms of the two-dimensional functions $u^{(1)},u^{(2)},u^{(3)},\ldots ,u^{(P)}$ respectively.  
In order to determine the Adomian matrix of Eq. \eqref{genF}, at first, we will start to determine the Adomian matrix for the first two matrices $U^{(1)},U^{(2)}$ or for the last two matrices $U^{(P-1)},U^{(P)}$ using the algorithms presented in the Step-1 to Step-6.
Let $A^{(P-1)(P)}$ is the Adomian matrix of the last two matrices $U^{(P-1)}$ and $U^{(P)}$. Next, we determine the Adomian matrix of the two matrices $A^{(P-1)(P)}$ and the previous one $U^{(P-2)}$. This process is continued up to first matrices $U^{(1)}$. After completing this process, finally, we will get the Adomian matrix of $F$ given in Eq. \eqref{genF}. We present this process in pseudo-code in Listing \ref{alg2} which determines the Adomian matrix of Eq. \eqref{genF}.

Now, we consider the nonlinear polynomial functional $F$ in the more general and complicated form (a sum raised to a power)
\begin{equation}\label{FpN}
	F=\left(u^{(1)}+u^{(2)}+u^{(3)}+\ldots+ u^{(P-2)}+u^{(P-1)}+u^{(P)}\right)^\mathcal{N}
\end{equation}
where the power index $\mathcal{N}$ is a positive integer number. In this case, at first, we expand Eq. \eqref{FpN} in sum of product terms. Then we can easily determine the Adomian matrix of each term of the expansion using the above algorithms for Eq. \eqref{genF}. Finally, simply adding all the Adomian matrices of each term, we get the Adomian matrix of Eq. \eqref{FpN}. 

In a one-dimensional case, the series \eqref{uvSeries} have only one index (say $i$). Therefore, all the matrices are one dimension, and in this case, in Step-3, we have to perform only a horizontal flipping operation. Besides this, all the algorithms described from Step-1 to Step-6 are identical in a one-dimensional case. In the following, we call the new algorithm presented by us the Adomian matrix algorithm.  

\section{Software implementation and comparisons with other algorithms}\label{comp}
We have implemented the algorithm described in Sec. \ref{algo} (called Adomian matrix algorithm) into MATHEMATICA code in Listings \ref{onedOur} (one-dimensional case), \ref{twodOur} (two-dimensional case) of Appendix: \ref{oned}, \ref{twod} respectively.
These MATHEMATICA programs can determine one-dimensional (using Listing \ref{onedOur}) and two-dimensional (using Listing \ref{twodOur}) Adomian polynomials of the following polynomial functional
\begin{equation}\label{FN}
	F=u^\mathcal{N},
\end{equation}
where the power index $\mathcal{N}$ is an positive integer number that represents the order of nonlinearity. To determine the Adomian polynomials of Eq. \eqref{FN}, we have to input the power index $\mathcal{N}$ and the order of the Adomian matrix in the function arguments (detailed descriptions of these function arguments are given in the Appendix) of the MATHEMATICA functions, and these functions print the Adomian polynomials in the output cell of the MATHEMATICA notebook. 

 MATHEMATICA codes for some other algorithms such as Duan’s Corollary 1 algorithm \cite{Duan}, Duan’s Corollary 3 algorithm \cite{Duan1} for one-dimensional case are also presented in Listings \ref{onedDuan}, \ref{onedDuan1} of Appendix: \ref{twod}.
 The MATHEMATICA programs in Listings \ref{onedDuan} and Listings \ref{onedDuan1} are taken from Appendix: A.1 in \cite{Duan} and from Appendix: A in \cite{Duan2} respectively. Here to make the programs more faster we have modified the programs (given in \cite{Duan}, \cite{Duan2}) which work only with the polynomial functional \eqref{FN} and evaluate the differentiation of \eqref{FN} using the factorial formula $\frac{d^iF}{du^i}=\frac{\mathcal{N}!}{(\mathcal{N}-i)!}u^{\mathcal{N}-i}$.
 
 We have compared the Adomian matrix algorithm with other algorithms by employing the MATHEMATICA programs given in Listings \ref{onedOur}, \ref{onedDuan}, \ref{onedDuan1}, \ref{twodOur} and using the polynomial functional \eqref{FN}.
In Table \ref{Tab:comp}, we have shown the comparisons between the computing speeds (measured in seconds) of the Adomian matrix algorithm (3rd column) and two different other algorithms (4th and 5th columns) for the one-dimensional case using the MATHEMATICA programs given in Listings \ref{onedOur}, \ref{onedDuan}, \ref{onedDuan1} in Appendix: \ref{oned}. We measure the computing times by MATHEMATICA
9.0 on the laptop with Intel(R) Core(TM) i5-7200U CPU $@$ 2.50 GHz and 8 GB RAM, using the MATHEMATICA command \verb+Timing[]+ with suppressing output (i.e., the results are retained in memory). 
Table \ref{Tab:comp} displays that the Adomian matrix algorithm is faster and more efficient than the other two algorithms: Duan’s Corollary 1 algorithm \cite{Duan} and Duan’s Corollary 3 algorithms \cite{Duan1}.
For example, we observe that in calculating the first $50$ Adomian polynomials, the Adomian matrix algorithm is almost $10^4$ times faster for $\mathcal{N}=3$ and almost $10^3$ times faster for $\mathcal{N}=10$ in comparison to the other two algorithms. Moreover, in calculating the first $100$ Adomian polynomials, the Adomian matrix algorithm spends the time $\sim 10^{-2}$ s, but, notably, the other two algorithms are unable to give results within an elapsed time of $600$ s.

We have also checked the computation efficiency of the Adomian matrix algorithm in the two-dimensional cases using the MATHEMATICA code in Listing \ref{twodOur}. For example, the Adomian polynomials of Eq. \eqref{FN} in the order of $40\times40$ are generated within $2.6$ s for $\mathcal{N}=3$ and within $19.5$ s for $\mathcal{N}=10$.

\begin{table}[H]
		\centering
		\caption{Comparisons of computing times (unit: seconds) of the Adomian matrix algorithm with some other algorithms using different values of $\mathcal{N}$ in \eqref{FN} and the different number $(n)$ of Adomian polynomials in one dimension. In some table cells, $\boldsymbol{\times}$ symbols indicate the algorithm in the corresponding column is unable to compute Adomian polynomials after spending almost $600$ s. }
		\begin{tabularx}{14.2cm}{|>{\centering\arraybackslash}p{2.1cm}|>{\centering\arraybackslash}p{2.9cm}|>{\centering\arraybackslash}p{2.4cm}|>{\centering\arraybackslash}p{3cm}|Y|}
			\hline
			 Nonlinearity \newline
			 $(\mathcal{N})$& Number of Adomian\newline polynomials $(n)$& Adomian matrix\newline algorithm&  Duan’s Corollary 1\newline algorithm \cite{Duan}& Duan’s Corollary 3\newline algorithm \cite{Duan1} \\
			\hline
			\multirow{4}{*}{3}&  10&  0.00047&  0.0020&  0.0025\\
			\cline{2-5}
			&  30&  0.002&  0.83&  0.76\\
			\cline{2-5}
			&  50&  0.0047&  62&  46\\
			\cline{2-5}
			&  100&  0.017&  $\boldsymbol{\times}$&  $\boldsymbol{\times}$\\
			\hline
			\multirow{4}{*}{5}&  10&  0.00078&  0.0026&  0.0025\\
			\cline{2-5}
			&  30&  0.0039&  0.87&  0.68\\
			\cline{2-5}
			&  50&  0.0092&  62.5&  46.4\\
			\cline{2-5}
			&  100&  0.037&  $\boldsymbol{\times}$&  $\boldsymbol{\times}$\\
			\hline
			\multirow{4}{*}{10}&  10&  0.0033&  0.0037&  0.0029\\
			\cline{2-5}
			&  30&  0.012&  0.96&  0.65\\
			\cline{2-5}
			&  50&  0.026&  62.7&  46.7\\
			\cline{2-5}
			&  100&  0.095&  $\boldsymbol{\times}$&  $\boldsymbol{\times}$\\
			\hline
		\end{tabularx}
	\label{Tab:comp}
	\end{table}

	\section{Conclusion}\label{con}
	We have presented a new algorithm (called the Adomian matrix algorithm) to determine the Adomian polynomials for scalar-valued nonlinear polynomial functional (with index as positive integers) in a Hilbert space $H$. The computations in the Adomian matrix algorithm do not need complicated mathematical operations such as parametrization, expansion, regrouping, differentiation, and so on. It is clear from Step-1 to Step-6 in Sec. \ref{algo} that the Adomian polynomials are determined entirely by some simple matrix operations. Because of the simplicity in mathematical operations, the algorithm is more efficient for the fast generation of the Adomian polynomials. We have designed two MATHEMATICA programs (one-dimensional case in Listing \ref{onedOur} and two-dimensional case in Listing \ref{twodOur}) based on the Adomian matrix algorithm, and compared its efficiency in computations for the one-dimensional cases with other two popular and powerful algorithms, which are Duan’s Corollary 1 algorithm \cite{Duan} and Duan’s Corollary 3 algorithms \cite{Duan1}. We have observed that the computation efficiency of the Adomian matrix algorithm is better than the other two algorithms. For example, in calculating the first $50$ Adomian polynomials in one dimension with the nonlinearity index $\mathcal{N}=3$ in Eq. \eqref{FN}, the Adomian matrix algorithm is almost $10^4$ times faster than the other two algorithms. 
	For $\mathcal{N}=10$, we are able to find the first $100$ Adomian polynomials using this new algorithm in just $10^{-2}$ s, whereas for $\mathcal{N}=3$ and $n=100$, the other two algorithms fail to produce any results until $600$ s have passed.
	Therefore, we can conclude that the Adomian matrix algorithm can be used to determine a large number of Adomian polynomials of nonlinear polynomial functionals that make the solutions more accurate.  
	
	\newpage
	\appendix
	
	\section{Mathematica programs for one-dimensional case }\label{oned}
	The following three MATHEMATICA programs can determine one-dimensional Adomian polynomials of the nonlinear function \eqref{FN}. The function arguments N\_ and n\_ represent the nonlinear power index $\mathcal{N}$ in Eq. \eqref{FN} and the number of first Adomian polynomials, respectively.
\begin{lstlisting}[language=Mathematica,keywordstyle=\color{blue}\bf,caption={Program based on the Adomian matrix algorithm.}, label={onedOur}]
  AdomMatAlgo1D[N_, n_] := Module[{h, j, k},
    u =.;
    mat = Table[Subscript[u, h], {h, 0, n - 1}];
    temmat = Table[Subscript[u, h], {h, 0, n - 1}];
    For[j = 1, j <= N - 1, j++,
      For[k = n, k >= 1, k--,
        mat[[k]] = Total[temmat[[;; k]]*Reverse[mat[[;; k]]]];
       ];
     ];
    mat
   ]
\end{lstlisting}
	
\begin{lstlisting}[language=Mathematica,keywordstyle=\color{blue}\bf,caption={Program based on the Duan's Corollary 1 algorithm \cite{Duan}.}, label={onedDuan}]
	DuanIndexAlgoAdom[N_, n_] := Module[{Apoly, Zpoly, dirClt}, 
	Subscript[Apoly, 0] = Subscript[u, 0]^N;
	Zpoly = Table[0, {i, 1, n - 1}, {j, 1, i}];
	Do[Zpoly[[suInd, 1]] = Subscript[u, suInd], {suInd, 1, n - 1}];
	For[i = 2, i <= n - 1, i++, 
	  For[j = 2, j <= i, j++, 
	    Zpoly[[i, j]] = Expand[Subscript[u, 1]*Zpoly[[i - 1, j - 1]]];
	    If[Head[Zpoly[[i, j]]] === Plus, 
	      Zpoly[[i, j]] = Map[#/Exponent[#, Subscript[u, 1]] &, Zpoly[[i, j]]], 
	      Zpoly[[i, j]] = Map[#/Exponent[#, Subscript[u, 1]] &, Zpoly[[i, j]], {0}]]];
	  For[j = 2, j <= Floor[i/2], j++, 
	    Zpoly[[i, j]] = Zpoly[[i, j]] + (Zpoly[[i - j, j]] /. 
	    Subscript[u, sub_] -> Subscript[u, sub + 1])]];
	dirClt = Table[Factorial[N]/Factorial[N - j]*(Subscript[u, 0]^(N - j)), {j, 1, n - 1}];
	Do[Subscript[Apoly, suInd] = Take[dirClt, suInd].Zpoly[[suInd]], {suInd, 1, n - 1}];
	Table[Subscript[Apoly, suInd], {suInd, 0, n - 1}]]
\end{lstlisting}
	
\begin{lstlisting}[language=Mathematica,keywordstyle=\color{blue}\bf,caption={Program based on the Duan's Corollary 3 algorithm \cite{Duan1,Duan2}.}, label={onedDuan1}]
   DuanCoro3AlgoAdm[N_, n_] := Module[{cPoly, i, k, j, derClt}, 
   Table[cPoly[i, k], {i, 1, n - 1}, {k, 1, i}];
   derClt = Table[Factorial[N]/Factorial[N - k]*(Subscript[u, 0]^(N - k)), 
                 {k,1, n - 1}];
   Apoly[0] = Subscript[u, 0]^N;
   For[i = 1, i <= n - 1, i++, 
     cPoly[i, 1] = Subscript[u, i];
     For[k = 2, k <= i, k++, 
       cPoly[i, k] = Expand[1/i*Sum[(j + 1)*Subscript[u, j + 1]*cPoly[i - 1 - j, k - 1], 
                           {j, 0,i - k}]]];
     Apoly[i] = Take[derClt, i].Table[cPoly[i, k], {k, 1, i}]];
   Table[Apoly[i], {i, 0, n - 1}]]
\end{lstlisting}
	
	\section{Mathematica programs for two-dimensional case }\label{twod}
	 The following MATHEMATICA program can determine two-dimensional Adomian polynomials of the nonlinear function \eqref{FN}.
	The function arguments N\_, m\_ and n\_ represent the nonlinear power index $\mathcal{N}$ in Eq. \eqref{FN}, the number of rows and number of columns in the Adomian matrix, respectively.
	
	\begin{lstlisting}[language=Mathematica,keywordstyle=\color{blue}\bf,caption={Program based on the Adomian matrix algorithm.}, label={twodOur}]
	  AdomMatAlgo2D[N_, m_, n_] := Module[{g, h, j, k, l},
	    u =.;
	    mat = Table[Subscript[u, g, h], {g, 0, m - 1}, {h, 0, n - 1}];
	    temmat = Table[Subscript[u, g, h], {g, 0, m - 1}, {h, 0, n - 1}];
	    For[j = 1, j <= N - 1, j++,
	      For[k = m, k >= 1, k--,
	        For[l = n, l >= 1, l--,
	          mat[[k, l]] = Total[temmat[[;; k, ;; l]]*Reverse[Reverse[mat[[;; k, ;; l]], 1], 2], 2];
	         ];
	       ];
	     ];
	    mat
	   ]
	\end{lstlisting}

\end{document}